%% file: main.tex
\documentclass[aps,pra,twocolumn,amsmath,amssymb,superscriptaddress]{revtex4-2}
\usepackage[utf8]{inputenc}
\usepackage[T1]{fontenc}
\usepackage{graphicx}  
\usepackage[dvipsnames]{xcolor}
\usepackage[colorlinks=true,linkcolor=blue,citecolor=red,urlcolor=magenta]{hyperref}
\usepackage{xspace}
\usepackage{siunitx}
\usepackage[version=4]{mhchem}
\usepackage{comment}
\usepackage{booktabs}
\usepackage{amsmath}

\newcommand{\aegis}{AE$\overline{\textrm{g}}$IS\xspace}
\newcommand{\pbar}{$\overline{\textrm{p}}$\xspace}

\begin{document} 

\title{Spectrometry of Captured Highly Charged Ions Produced Following Antiproton Annihilations}

\input{Authorlist} 
\collaboration{\aegis collaboration}
\begin{abstract}
We report a proof-of-principle study demonstrating the first capture and time-of-flight spectrometry of highly charged ions (HCIs) produced following antiproton annihilations in a Penning-Malmberg trap. A multi-step nested-trap technique was developed using the \aegis\ experiment to identify annihilation-linked captured ions. The trapping and spectrometry of helium and argon ions demonstrates the approach. This work establishes a foundation for the in-trap synthesis of radioactive HCIs and the study of cold nuclear annihilation fragments, with the long-term goal of enabling a sensitive tool for probing the outer nuclear periphery.
\end{abstract}

\maketitle

\textit{Introduction}.-- Precise knowledge of the spatial distributions of protons and neutrons in atomic nuclei is central to both nuclear physics and astrophysics. While electromagnetic probes can map charge densities with high precision \cite{Fricke2004,Nortershauser2020}, the neutron-rich peripheral region of nuclei remains experimentally challenging to explore. As an example, the neutron skin, given by the difference in neutron to proton root-mean-square radius, is a sensitive link to the nuclear forces~\cite{hu2022ab}, and its study allows us to explore the density dependence of the nuclear symmetry energy, which elucidates the properties of neutron stars~\cite{Brown2000_SkinSymm,Fattoyev2018}. The ongoing CREX-PREX dilemma~\cite{PREX2021,CREX2022,Reed2024,Kunjipurayil2025} further highlights the need for new techniques sensitive to the neutron distribution in the peripheral region of nuclei.

In this respect, the antiproton ($\bar{p}$) offers a unique probe, being notoriously sensitive to regions of low hadronic density as a result of its annihilation~\cite{Bugg1973,Richard2020}. With a mass approximately 1836 times that of the electron, while maintaining the same charge, allow the antiproton to be bound in a radially localized Rydberg orbit in close proximity to the atomic nucleus, as an antiprotonic atom~\cite{Bugg1973,backenstoss1989antiprotonic,JASTRZEBSKI1993405,doser2022antiprotonic,2025FPGStability}. The annihilation on either a proton or neutron, results in the release of roughly $\sim$1.5~GeV captured by multiple mesons ejected from the annihilation site~\cite{amsler2024antiproton}. If the annihilation occurs in the far periphery then the mesons may miss the residual nucleus, leaving behind a 'cold' fragment with minimal recoil momentum and one less nucleon~\cite{JASTRZEBSKI1993405,Wycech1996}. The ratio of neutron ($n$) to proton ($p$) loss of these cold fragments mirrors the local ($n/p$) density ratio in an outer peripheral region $\sim$3~fm beyond the nuclear surface, where the nuclear density is up to $10^3$ times lower than the core~\cite{Polster1995,Schmidt1999}. The study of these cold fragments offers a sensitive constraint on the neutron skin and the isospin structure of the extreme peripheral region of the nucleus~\cite{Trzcinska2001}.

Historically, the study of antiproton interactions with matter has been performed via beam-on-target (gas, liquid, or solid) experiments \cite{backenstoss1989antiprotonic,widmann_phase_1995}, which has provided valuable data on antiproton ionization and capture cross-sections for benchmarking atomic theory \cite{andersen_single_1986,bacher_degree_1988,cohen_capture_2004-1,kirchner2011current,PhysRevA.110.012803}. Following the antiprotonic atom cascade and annihilation, X-rays, high energy mesons and light, energetic annihilation fragments (such as neutrons, protons, deuterons, tritons and helium nuclei) have all been directly detected \cite{Bugg1973,Lubi2002,2008GottaEPJD,MARKIEL1988445,von1990yield,Chen1992,amsler2024antiproton,Berghold2025}. However, the much heavier nuclear fragments have only been studied using radiochemical means through gamma spectroscopy \cite{Polster1995,Lubi2002}. Due to the limitations of this radiochemical approach, the cold fragment ratios could only be deduced for selected cases with appropriate decay properties \cite{JASTRZEBSKI1993405,lubinski1994neutron,Lubin1998,Schmidt1999}. These measurements allowed systematic studies of the neutron skin by constraining models together with X-ray spectroscopy measurements of antiprotonic atoms and electron scattering data \cite{Trzcinska2001,Schmidt2003}. However, cases involving non-radioactive cold fragments, such as those from $^{208}$Pb, had to be interpolated \cite{Klos2007}, introducing a significant systematic uncertainty for the determination of their neutron skin.

As a result of the technical advancements at the AD-ELENA facility at CERN, the cooling, trapping and accumulation of millions of antiprotons in Penning-Malmberg traps is now routine \cite{carli2022elena,caravita2025cernadelenaantimatterprogram}. This has opened the possibilities for studies of antiprotonic atoms formed in a controlled Ultra-High-Vacuum (UHV) environment, as first proposed by Wada and Yamazaki~\cite{wada2004technical}. More recently, the PUMA experiment at CERN is aiming at probing the nuclear periphery of short-lived radioactive nuclei using trapped antiprotons, by measuring the net charge of mesons from the annihilation. This method explores a region $\sim$2-2.5~fm from the nuclear half-density radius~\cite{aumann2022puma}. A complementary approach, conceptually outlined here, aims at the direct capture and time-of-flight (TOF) spectrometry of recoil-filtered cold annihilation fragments in a trap. This technique allows for the in-situ trapping of charged species resulting from the antiproton annihilation, including both nuclear annihilation fragments and ions produced via the ionization of the trap environment. Specifically, the envisioned mass spectrometry of the captured cold fragments will enable a novel tool for probing the outer peripheral region $\sim$3~fm beyond the half-density radius, while filtering events where high energy mesons strongly interact with the residual nucleus, by the cold fragment recoil energy. Antiproton-based probes of the neutron periphery are known to involve sizable model dependence~\cite{Wycech1996}, including assumptions about the annihilation mechanism, fragment survival, and bound-state antiproton overlap with the nuclear matter distribution. This approach could help constrain these model dependencies through simultaneous charge and mass measurements of recoil-filtered annihilation fragments. The high charge-state of annihilation fragments, as a result of the Meitner-Auger cascade, enables their efficient capture within a Penning-Malmberg trap, as indicated by recent simulations~\cite{kornakov2023synthesis}. Furthermore, these often radioactive highly charged ions (HCIs), if captured, are valuable systems for precision studies of fundamental interactions and metrology~\cite{morgner2023stringent,schweiger2024penning,Micke2020,blaum2020perspectives,kimura2023hyperfine,Sara2025}.

In this work, we report the experimental validation of this novel technique via a first proof-of-principle study, demonstrating the capture of annihilation-induced ions, a crucial step for the future capture of nuclear annihilation fragments. We demonstrate the procedure by capturing highly charged ions produced from antiproton annihilations occurring on ultra-low-density helium and argon gas within a Penning-Malmberg trap. Using the AEgIS experiment at CERN~\cite{Amsler2021}, we employed TOF spectrometry to identify the captured ion species. These first results establish the core methodology for producing and trapping radioactive HCIs and outline a path towards the high-resolution spectrometry of cold annihilation fragments, which will give access to new observables to study the antiproton annihilation process, and serve as a powerful and complementary tool to probe the outer nuclear periphery.

\begin{figure*}[tb]
  \centering
  \includegraphics[width=0.8\textwidth]{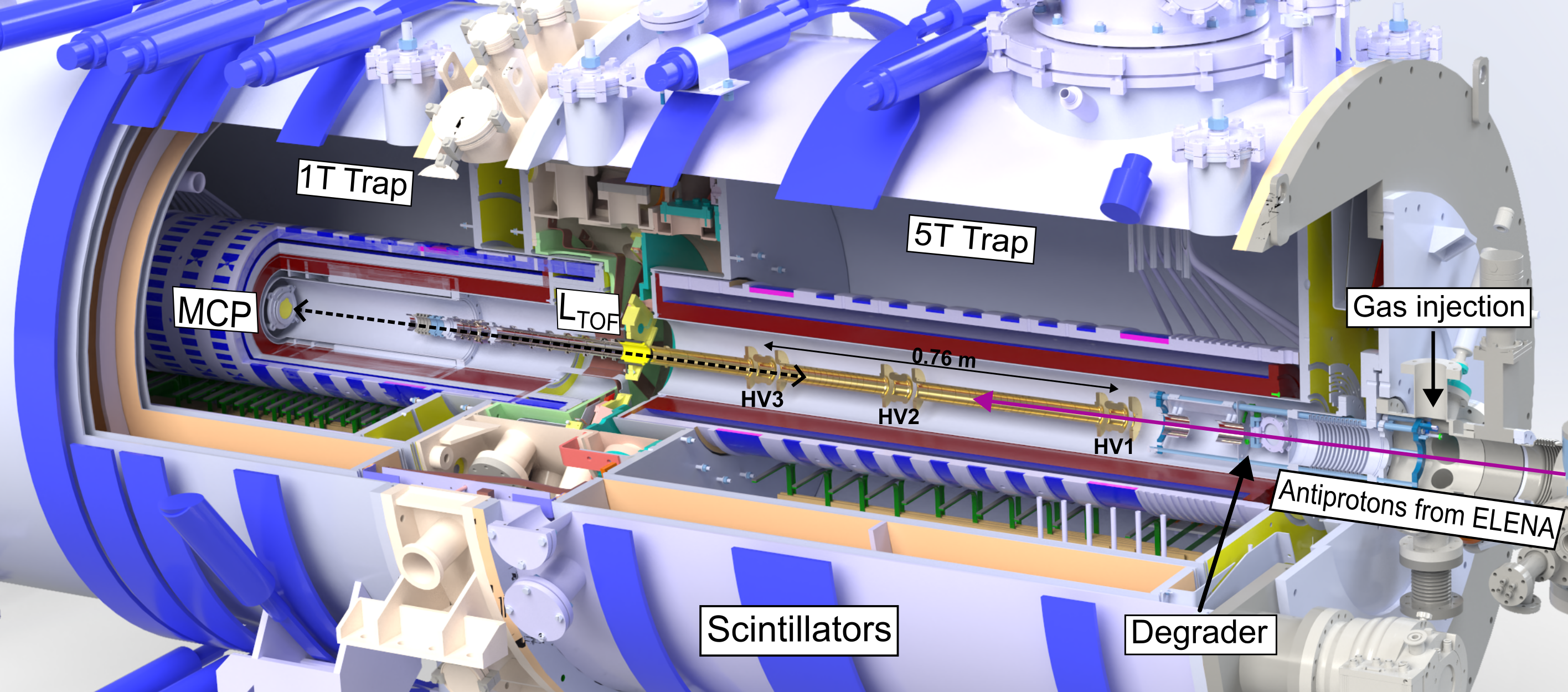}
  \caption{Overview of the \aegis\ experimental setup. Antiprotons decelerated by the AD-ELENA decelerators enter the \aegis\ \qty{5}{\tesla} trap after passing a degrader foil. The antiprotons are trapped in the \qty{5}{\tesla} trapping region, between the HV1 and HV3 electrode biased to $V_{HV}=$\qty{-14}{kV}. Buffer gas was injected using a leak valve at the entrance of the \qty{5}{\tesla} trapping region. Annihilations occurring within the trapping region are detected via ejected mesons using external scintillators shown in blue. Positive ions are captured using a nested trap potential presented in Fig. \ref{fig:procedure} and ejected axially to a downstream MCP, placed at a distance of $L_{\text{TOF}}=1.05(1)$m within the 1T trapping region for Time-of-Flight (TOF) analysis of the charged species.}
  \label{fig:AEGISexperiment}
\end{figure*}

\begin{figure}[h]
\center
\includegraphics[width=0.9\columnwidth]{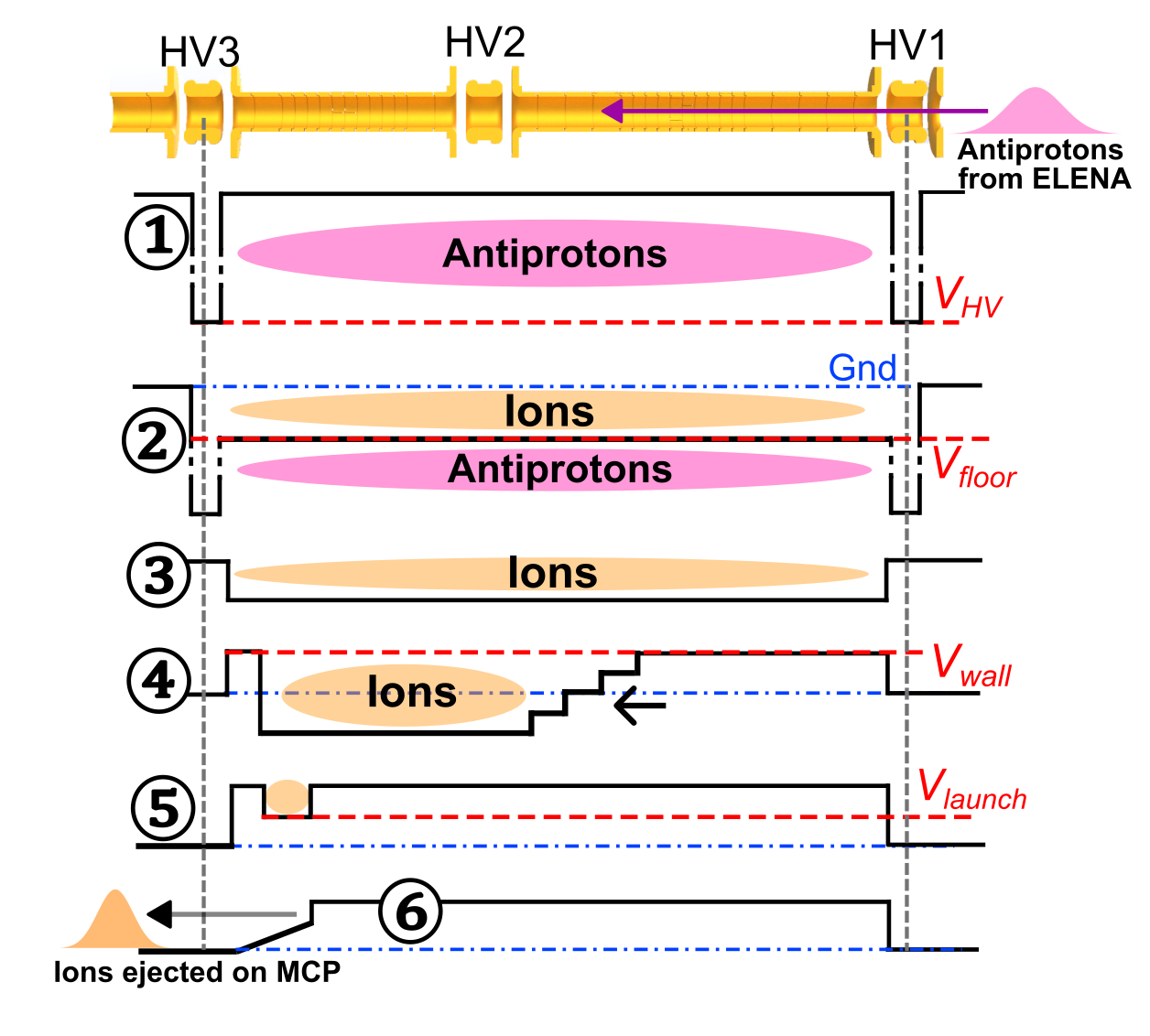}
\caption{Overview of the multi-step nested trap procedure, used for the capture and TOF spectrometry of positive ion species formed following antiproton annihilations in the \qty{5}{\tesla} \aegis\ Penning-Malmberg trap. Each step is presented in chronological order from 1 to 6, showing the applied voltages on the trap electrodes together with the antiproton and ion location within the trap. The \qty{0}{V} potential of the trap is indicated with a blue dash-dotted line, the red dashed lines indicates relevant voltage settings.}
\label{fig:procedure}
\end{figure}

\textit{Experiment}.-- The experiment was conducted in the AD-ELENA antiproton decelerator complex at CERN~\cite{carli2022elena}, using the \aegis\ Penning-Malmberg traps, seen with cross-sectional view in Fig. \ref{fig:AEGISexperiment}. Antiproton bunches containing $\sim8 \times 10^6$ antiprotons per bunch at \qty{100}{\keV} delivered from ELENA, were injected into the cryogenic Penning-Malmberg trap (<\qty{10}{K}) within the \qty{5}{\tesla} superconducting solenoid after passing a thin degrader foil, reducing the kinetic energy to below $\sim$\qty{20}{\keV} for the dynamic capture between high-voltage electrodes (HV1 and HV3). A multi-step nested-trap procedure, seen in Fig. \ref{fig:procedure}, was developed using the Penning-Malmberg trap electrodes within the \qty{5}{\tesla} trapping region, to capture and eject trapped ions for TOF spectrometry. Following $\bar{p}$ capture using $\mathrm{V_{HV}}=$~-14~kV (step 1), a nested potential well was formed by biasing multiple low-voltage electrodes (range of $\pm$200 V) to $V_{\text{floor}}$, enabling the capture of positive ions produced from $\bar{p}$ interactions. Charged pions emitted from annihilation events were simultaneously monitored via external scintillators, shown in blue in Fig. \ref{fig:AEGISexperiment}. After a set antiproton trapping time ($\tau_{\overline{p}}$), the remaining $\bar{p}$ were ejected by grounding the HV1 and HV3 electrodes (step 3). Trapped positive ions were subsequently compressed axially (step 4) and prepared for TOF analysis by raising the trap floor to $\mathrm{V_{launch}}$ with walls at $\mathrm{V_{wall}}$ (step 5), effectively filtering hot ions before release. The remaining ions in the shallow trap were then ejected to a downstream Micro-Channel Plate (MCP) detector located $L_{\text{TOF}}=1.05(1)\text{ m}$ downstream located within a \qty{1}{\tesla} superconducting solenoid (step 6). The complete procedure was performed under good vacuum conditions (< $10^{-15}$ mbar), with and without trapped antiprotons to characterize any background signal. The TOF spectrometry step of the procedure was calibration using electron-cooled antiprotons captured using the same electrodes but with a reverse polarity trap (step 5 and 6). 
The complete experimental procedure was tested with the controlled injection of ultra-low-density gases at the entrance of the \qty{5}{\tesla} solenoid, first using helium followed by a mixture of helium and argon gas while maintaining a gauge pressure reading between $10^{-7}$ to $10^{-8}\text{ mbar}$, at the entrance of the \qty{5}{\tesla} superconducting solenoid. A detailed description of the experimental procedure is given in the Supplemental Materials.

\begin{figure*}[tb]
  \centering
  \includegraphics[width=1\textwidth]{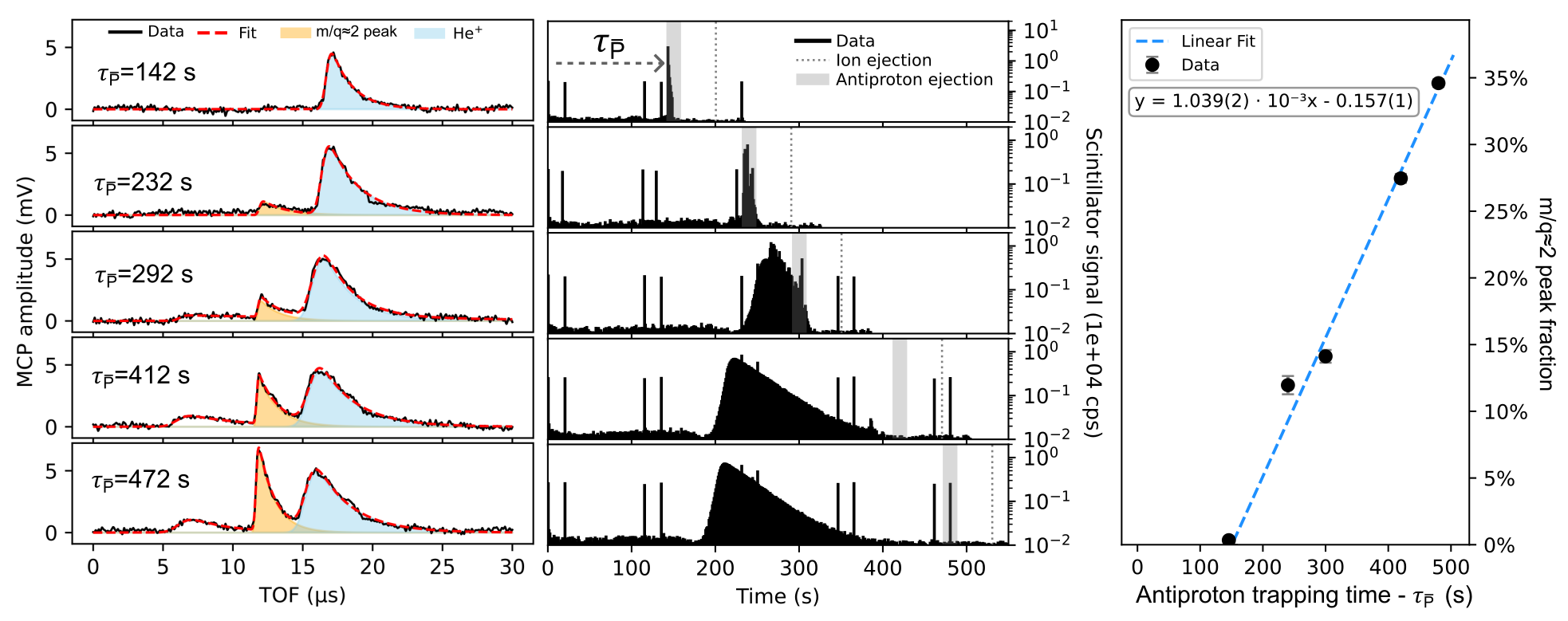}
  \caption{TOF measurements in Helium background, showing MCP signal observed at different antiproton storage times ($\tau_{\overline{\mathrm{p}}}$) together with measured annihilation events using external scintillators. Left and center panels: MCP signal vs TOF (left panel) together with corresponding scintillator signal vs time after antiproton capture (center panels). The MCP signal shows peaks identified as \ce{He^1+} (m/q $\approx$ 4) and (m/q $\approx$ 2) species, the center panels shows the scintillator events observed for each measurement, the grey shade shows the release of antiprotons, and the dotted line indicate the ion ejection time. The repeated spikes observed in the scintillator signal are from the AD-cycle background. Right panel: plot of the integrated fractional intensity of the m/q $\approx$ 2 peak with respect to $\tau_{\overline{\mathrm{p}}}$, fitted with the linear function shown in the inset.}
  \label{fig:evolution}
\end{figure*}

\textit{Results}.-- The result of the multi-step nested trap approach (using a nested trap with the settings: $\mathrm{V_{floor}}=$~-160~V, $\mathrm{V_{launch}}=$~90~V and $\mathrm{V_{wall}}=$~160~V) tested with helium injection are presented in Fig.~\ref{fig:evolution}, revealing the evolution of the TOF spectrum (left panels) and scintillation signal (center panels) with respect to different antiproton trapping times ($\tau_{\overline{\mathrm{p}}}$). The antiproton ejection time is indicated with a grey shade for each measurement, with a dotted vertical line indicating the ion ejection time. When antiprotons are released from the trapping region before annihilations are detected, only a single ion peak is observed in the spectrum. As shown in the top-left panel of Fig. \ref{fig:evolution}, this peak at \qty{16.5(4)}{\micro\s} corresponds to a species with a mass-to-charge ratio of $m/q$=3.8(2) u/e according to the antiproton calibration, which is in agreement with expected He$^{+}$ ions.

The average resolution achieved in these measurements was $M/\Delta M \approx 4$, significantly lower than for the electron-cooled antiprotons which achieved a mass resolution of $M/\Delta M \approx 45$, details given in Supplemental Materials. As the antiproton trapping time is increased, annihilations are observed inside the trapping region during the ion capture sequence, center panels in Fig. \ref{fig:evolution}. As a result, a second peak appears at \qty{12.0(2)}{\micro\s} corresponding to ion species with $m/q$=2.01(6) u/e suggesting the capture of either \ce{H2^+}, \ce{He^2+} or HCI fragments, all with $m/q\approx$2 u/e. Background measurement under good vacuum conditions indicated no detectable MCP signal. A third peak appears around \qty{6}{\micro\s} which is attributed to ions not sufficiently cooled before release, which was confirmed by stepwise reducing the trap depth by lowering $\mathrm{V_{wall}}$. The fractional peak intensity of the $m/q\approx$2 u/e peak vs the antiproton trapping time is seen in the right panel of Fig.~\ref{fig:evolution}. A linear increase of the $m/q\approx$2 u/e ion fraction is seen, directly following the observation of annihilations in the trapping region.

\begin{figure}[h]
\center
\includegraphics[width=1\columnwidth]{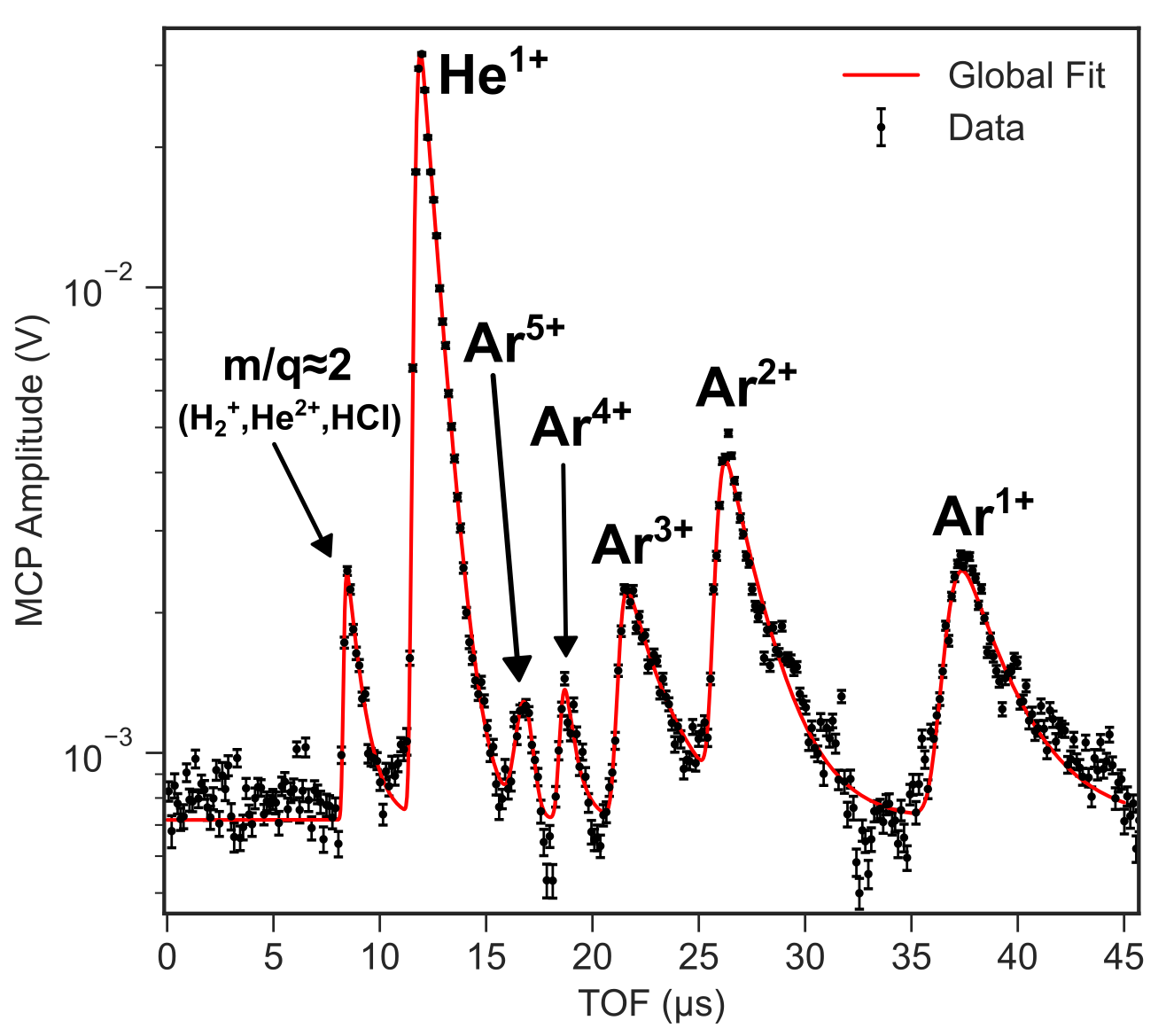}
\caption{TOF spectrum of ejected positive ions observed during the argon/helium  experiment. Data (points) are shown with the multi-peak fit (red line), further details are provided in the Supplemental Materials.}
\label{Fig:Fitted_spectrum}
\end{figure}

In the following experiment, argon was introduced together with a helium background. Using a nested trap configuration with $\mathrm{V_{wall}}=$~-190 V, $\mathrm{V_{launch}}=$~180 V and $\mathrm{V_{wall}}=$~190 V. A shallower trap was used to reduce the energy spread of the ion bunch through evaporative cooling, filtering the ions with energies above $\sim$5~eV. Fig.~\ref{Fig:Fitted_spectrum} shows the acquired TOF spectrum, representing the average sum of four measurements, captured following the complete annihilation of all antiprotons in the trapping region after $\sim$\qty{150}{\s} trapping time. The fit results are overlaid on the spectrum which exhibit seven distinct peaks, which are presented on a log-lin scale to enhance clarity for the weaker signals. Tab.~\ref{Fig:tableAr} summarizes the identified peaks, their measured TOF, and calculated $m/q$ values using the \ce{He^1+} peak as a calibration reference. In these measurements an average mass resolving power of $M/\Delta M \approx 5$ was achieved, sufficient to distinguish the different charge states in the spectrum. Besides the \ce{He^1+} peak and the signal at $m/q$ $\approx$ 2 u/e, several new peaks appear at $m/q$ values consistent with argon ions in various charge states, ranging from \ce{Ar^1+} through \ce{Ar^5+}. Higher charge states could not be identified due to the large \ce{He^1+} signal. The TOF values are in reasonable agreement with the literature values shown, although a slight systematic deviation towards higher $m/q$ is observed for the heavier ions, see Supplemental Materials.

\begin{table}[h]
    \centering
    \setlength{\tabcolsep}{6pt}
    \renewcommand{\arraystretch}{1.3}
    \caption{Identified ion peaks from the Argon campaign spectrum (Fig.~\ref{Fig:Fitted_spectrum}). The spectrum was calibrated using TOF of the identified \ce{He^+} peak, \(m/q\) values are given with statistical uncertainties (parentheses). The literature values are taken from  Ref.  \cite{Huang_2021}.}
    \begin{tabular}{cccc}
    \toprule
    \textbf{TOF (\si{\micro\second})} & $\frac{m}{q}\big|_{\text{exp}}$ \(\left(\frac{\mathbf{u}}{\mathbf{e}}\right)\) & \textbf{Ion Candidate(s)} & $\frac{\mathbf{m}}{\mathbf{q}}\big|_{\textbf{lit}}$ \(\left(\frac{\mathbf{u}}{\mathbf{e}}\right)\) \\
    \midrule
    \( 8.457(30) \)  & \( 2.018(14) \)  & 
    \begin{tabular}{@{}c@{}} \ce{H2^+} \\ \ce{He^2+} \\ HCI frag.\footnote{Assuming the complete stripping of nuclear annihilation fragments with masses greater than $^4$He.} \end{tabular} &
    \begin{tabular}{@{}c@{}} 2.01510 \\ 2.00075 \\ 2.0-2.6 \end{tabular} \\
    \( 11.911(7) \)  & --              & \ce{He^1+}            & 4.00260 \\
    \( 16.774(75) \) & \( 7.94(7) \)   & \ce{Ar^5+}            & 7.99193 \\
    \( 18.698(68) \) & \( 9.86(7) \)   & \ce{Ar^4+}            & 9.99005 \\
    \( 21.641(56) \) & \( 13.21(7) \)  & \ce{Ar^3+}            & 13.32025 \\
    \( 26.264(29) \) & \( 19.46(5) \)  & \ce{Ar^2+}            & 19.98064 \\
    \( 37.378(55) \) & \( 39.41(12) \) & \ce{Ar^1+}            & 39.96183 \\
    \bottomrule
    \end{tabular}
    \label{Fig:tableAr}
\end{table}

\textit{Discussion}.-- The presented measurements establish a first demonstration of the capture, cooling and TOF spectrometry of ions formed following antiproton annihilations in a Penning–Malmberg trap. The demonstrated simultaneous measurement by scintillator tagging with ion yields also shows that external scintillators provide a practical handle to identify annihilation-linked ion production.

In helium buffer gas, the nearly time-independent \ce{He+} yield is consistent with the direct collisional ionization during \pbar\-injection, whereas the growth of the \(m/q\!\approx\!2\) u/e peak is a result of a secondary process following the antiproton annihilations occurring within the trapping region, suggesting annihilation-linked ion production. Direct capture of $\overline{\textrm{p}} ^{4}$He fragments is excluded, as their recoil energy is far beyond the potential depth of the trap. Antiproton annihilations occurring on heavier elements such as the gold plated electrodes are expected to result in a large distribution of highly charged nuclear fragments~\cite{Lubi2002}. While a small contribution from directly trapped primary nuclear fragments, especially for the heavier \ce{Ar} target, cannot be excluded, the current charge-state distribution, together with the absence of an efficient and charge state preserving cooling mechanism for fast HCI fragments favors secondary ionization as the dominant origin under these conditions. A more likely scenario is that the HCI annihilation fragments or energetic electrons impact ionize the residual gas within the trapping region. Reference data for slow $\overline{\textrm{p}}$-He and Ar collisions show predominantly \ce{He+} and \ce{Ar^{2+}} up to \(\sim\!25~\text{keV}\) impact energy~\cite{knudsen2008ionization}. Thus the observation of \ce{Ar^{5+}}, which requires a cumulative ionization energy exceeding \qty{200}{eV}, at \pbar\ energies \(<\!14~\text{keV}\) points to the presence of highly ionizing particles within the trap volume, that drive secondary ionization of the buffer gas, which could generate \ce{He^{2+}} ions. In this scenario the intensity is expected to decrease with increasing charge state, notably $\text{Ar}^{1+}$ showed a lower yield than $\text{Ar}^{2+}$. This is explained by $m/q$ normalization and losses of the low $m/q$ ions by the magnetic bottleneck ($\text{5 T}$ to $\text{1 T}$). We note that Hydrogen is a well-known impurity in Penning-Malmberg traps, even under UHV conditions~\cite{OKADA2023128617}, \ce{H2^+} ions could thus also be a possible candidate. However, background measurements running the sequence with He/no-\pbar and no-He/\pbar indicated no detectable MCP signal, supporting \ce{He^{2+}} ions as the most likely candidate. A refined technique with improved mass resolution could easily isolate any contamination during future measurements.

The present buffer-gas environment, used in this work simultaneously as a target for the ion formation and cooling of the species, limits sensitivity to the most interesting stripped annihilation fragments, HCIs are depleted by charge exchange, and primary fragments may scatter before capture. A direct path forward, to directly isolate the primary nuclear fragments, is to perform measurements without the presence of buffer gas to preserve fragment charge states and kinematics. Refined approaches utilizing \pbar$-$anion mixing would allow the laser-triggered synthesis of antiprotonic atoms and fragments in UHV~\cite{Gerber2019,kornakov2023synthesis}, and for species not forming anions, a Rydberg excited atomic beam could be injected on the trapped cold antiproton target, similar to positronium formation~\cite{Hessels1998}.

The experiments reported here were constrained by a \(\sim\)\qty{190}{\volt} well depth set by available voltage supplies and electrode geometries. Furthermore, captured ions were not sufficiently cooled before their ejection from the trap for TOF spectrometry. Enabling positive bias on the HV electrodes will substantially increase capture efficiency of charged species to energies beyond \(\qty{10}{\kilo\electronvolt}/q\). This will allow the efficient capture of cold annihilation fragments for nuclear periphery studies, with the estimated complete capture of cold fragments from annihilation on $^{80}$Kr and heavier nuclei~\cite{kornakov2023synthesis}. We note that the extraction of nuclear structure information from these measurements relies on theoretical descriptions of the antiproton–nucleon/nucleus interactions, which motivate the need for modern theoretical approaches~\cite{Schmidt2024,Pachucki2025,sommerfeldt2025} to fully interpret future data.

Achieving the mass resolution required for isotopic and isobaric separation will require efficient, non-intrusive cooling of the captured HCIs. Promising options include combined electron/positron cooling~\cite{OSHIMA2003,OSHIMA2005} and sympathetic cooling with laser-cooled ions~\cite{BUSSMANN2006}. Once the fragments are cooled to thermal energies, a TOF resolving power comparable to that obtained with electron-cooled antiprotons (\(M/\Delta M\!\sim\!100\)) should be reachable at \aegis, using the demonstrated technique. For example, to identify $A-1$ fragments of fully stripped $^{80}$Kr fragments following annihilation, a minimal mass resolution above \(M/\Delta M\!>\!35\) is necessary. To resolve multiple charge states of different isotopes and isobars, multi-reflection TOF (MR-TOF) mass spectrometry is a promising approach which can reach resolving powers of \(M/\Delta M\!>\!10^5\)~\cite{VERENCHIKOV2025117395,maier2025highvoltagemrtofmassspectrometer}. As a complementary tool, mass-selective RF techniques could provide in-trap filtration of specific \(m/q\) bands~\cite{Huang1997}. 

Once implemented, this technique will enable systematic studies of charge-state distributions across different target atoms, thereby constraining antiproton collisional-ionization cross-sections and the capture/cascade process, valuable for exotic atom theory~\cite{bacher_degree_1988,cohen_capture_2004-1,Bailey2015,Jonsell2018,PhysRevA.110.012803}. 
Single-nucleon-loss fragment ratios can map the changes in the neutron to proton density at the nuclear periphery along isotopic chains, providing a powerful complement to nuclear charge-radius data~\cite{Nortershauser2020,Gustafsson2025}. These measurements could reveal subtle changes in the neutron skin, constraining the nuclear forces~\cite{hu2022ab} and the Equation-of-State of neutron stars~\cite{Brown2000_SkinSymm,Fattoyev2018}. This unique probe could potentially also be sensitive to surface $\alpha$-clustering in nuclei~\cite{Junki2021,Cox2025Te104}. In addition, measurements of A-2 and higher-order fragment, if captured, could yield valuable benchmarks for intranuclear cascade models \cite{kornakov2023synthesis,amsler2024antiproton} and multi-nucleon annihilation processes, such as Pontecorvo reactions \cite{Chiba1997,Venturelli:2025ES}.

\textit{Conclusions}.-- This proof-of-principle work establishes a practical approach for studying annihilation-induced ion production within a Penning-Malmberg trap, demonstrating the essential steps for direct capture and TOF spectrometry of nuclear annihilation fragments. Technical developments, such as laser-triggered annihilation in UHV, deeper nested wells, non-intrusive cooling, and high-resolution mass analysis (MR-TOF and/or mass-selective RF), are readily achievable. Once implemented, studies of the charge and mass of recoil-filtered nuclear annihilation fragments will offer unique insight into the life-cycle of antiprotonic atoms, while enabling a powerful complementary tool to probe the outer nuclear periphery, with broad implications for atomic, nuclear, particle and astrophysics. The future availability of transportable antimatter traps~\cite{aumann2022puma,leonhardt2025proton} together with the presented approach could extend this technique to probe the outer nuclear periphery of short-lived nuclei and allow an alternative path for synthesizing radioactive HCIs in facilities without access to large accelerator infrastructure.

\begin{acknowledgments}
We wish to express our thanks to Eric Vidal Marcos, Naofumi Kuroda and Alexandre Obertelli for valuable discussions and suggestions. We thank B. Bergmann, P. Burian, S. Pospisil, P. Smolyanskiy and V. Petracek for financial support. This work was supported by the Istituto Nazionale di Fisica Nucleare; the CERN Fellowship programme and the CERN Doctoral student programme; the EPSRC of UK under grant number EP/X014851/1; the Research Council of Norway under Grant Agreement No. 303337 and NorCC; the CERN-NTNU doctoral program; the European Union’s Horizon Europe research and innovation program under the Marie Skłodowska-Curie grant agreement Nr 101109574; the Research University – Excellence Initiative of Warsaw University of Technology via the strategic funds of the Priority Research Centre of High Energy Physics and Experimental Techniques; the IDUB POSTDOC programme; the IDUB YoungPW programme under agreement no. IDUB/56/Z01/2024; the Polish National Science Centre under agreements no. 2022/45/B/ST2/02029,2022/46/E/ST2/00255 and 2023/50/E/ST2/00574, and by the Polish Ministry of Education and Science under agreement no. 2022/WK/06; the Marie Sklodowska-Curie Innovative Training Network Fellowship of the European Commission's Horizon 2020 Programme (No. 721559 AVA); the European Union's Horizon 2020 research and innovation programme under the Marie Sklodowska-Curie grant agreement ANGRAM No. 748826; the Wolfgang Gentner Programme of the German Federal Ministry of Education and Research (grant no. 13E18CHA); the European Research Council under the European Union's Seventh Framework Program FP7/2007-2013 (Grants Nos. 291242 and 277762); and the European Social Fund within the framework of realizing the project, in support of intersectoral mobility and quality enhancement of research teams at Czech Technical University in Prague (Grant No. CZ.1.07/2.3.00/30.0034).
\end{acknowledgments}

\section*{Supplemental Materials}

\subsection{Detailed Experimental Methodology}

The \aegis\ experiment is composed of two superconducting solenoid magnets with two connected Penning-Malmberg trapping regions, a \qty{5}{\tesla} region used for catching and preparing antiprotons, and an adjacent \qty{1}{\tesla} trapping region used for antihydrogen production. In these studies, only the Penning-Malmberg trap electrodes between the High-Voltage HV1 and HV3 electrodes were used, all the trap electrodes in the \qty{1}{\tesla} region were grounded. Bunches of approximately $8 \times 10^6$ antiprotons at \qty{100}{\keV} kinetic energy were delivered from the AD-ELENA decelerator complex every $\sim$\qty{110}{\s} \cite{gamba2021ad}. These \pbar\ bunches impinged on a \qty{1400}{\nm}-thick Mylar degrader foil held on a flippable mount positioned inside a \qty{5}{\tesla} superconducting solenoid magnet. This degrader reduces the \pbar\ energy sufficiently (below $\sim$\qty{20}{\keV}) for dynamic capture within the \qty{5}{\tesla} Penning-Malmberg trap. Antiproton capture was achieved by rapidly pulsing (\qty{10}{ns} scale) the high-voltage (HV1) electrode, to potentials up to \qty{-14}{\kV} with the HV3 maintained at \qty{-14}{\kV}. Typically, around $3 \times 10^6$ antiprotons were captured per cycle. Between HV1 and HV3 electrodes reside 20 cylindrical ring electrodes (\qty{3}{cm} diameter and 1.5-\qty{3}{cm} long) made from gold plated copper, including the high-voltage electrode HV2, which can be individually biased up to \qty{\pm 200}{\volt} (\qty{10}{\mV} resolution) via the \aegis\ control system \cite{Volponi2024,volponiTALOSTotalAutomation2024}, allowing for precise manipulation of trapped charged particles. The trap electrodes are cooled to cryogenic temperatures ($\sim$ \qty{10}{K}). Controlled injection of either helium or argon gas (\SI{99.9}{\percent} purity) into the \aegis\ trapping region was achieved using an adjustable high-vacuum leak valve connected to a reservoir tuned to a pressure between \SIrange{10}{100}{\milli\bar}. The leak valve was regulated to maintain a pressure reading of \SIrange{e-7}{e-8}{\milli\bar} at a vacuum gauge located at the entrance to the \SI{5}{\tesla} trapping region, for the duration of the experimental procedures. The pressure profile within the trapping volume itself is expected to deviate substantially from this upstream value due to the strong cryogenic pumping by the cold bore and electrodes. To prevent back-streaming of the injected gas and potential contamination of ELENA, an upstream gate valve was closed and actuated only for brief periods (\SI{\sim 5}{\second}) synchronized with the injection of the antiproton beam into the \aegis\ apparatus. The antiproton degrader was opened throughout the experimental procedure except when needed during the antiproton injection (lasting \SI{\sim 10}{\second}). The inefficient cryopumping of helium resulted in a contamination present during the argon injection studies.

The experimental cycle proceeded as follows for every captured antiproton bunch arriving from ELENA into the \aegis\ trapping region (see Fig.~\ref{fig:procedure}).

\begin{enumerate}
    \item[1] Antiprotons are captured between the pulsed high-voltage ($V_{HV}=-14$ kV) electrodes. Collisions with the buffer gas cool the antiprotons, resulting in the ionization of the atoms. 
    \item[2] The low-voltage electrodes between HV1 and HV3 were then biased to a potential $V_{\text{floor}}$ (set to \qty{-160}{\volt} for helium and \qty{-190}{\volt} for the argon/helium measurements), forming a nested potential well that allowed for the capture of any positive species. Annihilations occurring within the trapping region are monitored using external scintillators that detect pions released during annihilation.    
    \item[3] After a set antiproton trapping time ($\tau_{\overline{\mathrm{p}}}$), the potential of the HV3 electrode is ramped down to ground over approximately \qty{12}{\s}. This process ejects the remaining antiprotons axially. These ejection-induced annihilations are monitored using the external scintillators, providing a measure of the remaining antiproton number. Positive ions with insufficient axial energy to overcome the potential $V_{\text{floor}}$ remain trapped in the central region.
    \item[4] The potential on the low voltage electrodes is manipulated over approximately \qty{10}{\s} to axially compress the captured positive ions into a shorter trap (approximately \qty{4}{\cm} long) formed by three adjacent 1.5 cm long electrodes near the downstream HV3 electrode.
    \item[5] The potential of this short trap is raised over approximately \qty{5}{\s} to form a harmonic well with a set floor potential $\mathrm{V_{launch}}$ (\qty{+90}{\volt} for helium and \qty{+180}{\volt} for argon), confined axially by wall potentials $\mathrm{V_{wall}}$ ( \qty{+160}{\volt} for helium and \qty{+190}{\volt} for argon). The launch potential $\mathrm{V_{launch}}$ defines the potential energy of the ions ejected for the TOF measurement. This ion manipulation sequence took approximately \qty{20}{\s}.
    \item[6] Finally, the downstream wall electrode is rapidly switched (\qty{1}{\ns} timescale) to \qty{0}{\volt}, ejecting the trapped positive ions axially towards a micro-channel plate (MCP) detector located at a distance of $L_{\text{TOF}}$=\qty{1.05(1)}{\m} downstream within the \qty{1}{\tesla} magnetic field of the trapping region. The front face of the MCP was biased to \qty{0}{\volt}. The MCP readout is amplified and recorded using a Teledyne LeCroy HDO6104 oscilloscope, triggered synchronously with the ejection pulse, yielding the captured TOF signal. The scintillator signals are processed using counting modules to monitor annihilation rates and timing throughout each experimental cycle.
\end{enumerate}

\subsection{Time-of-Flight fitting Procedure}

\begin{figure}[h]
\center
\includegraphics[width=1\columnwidth]{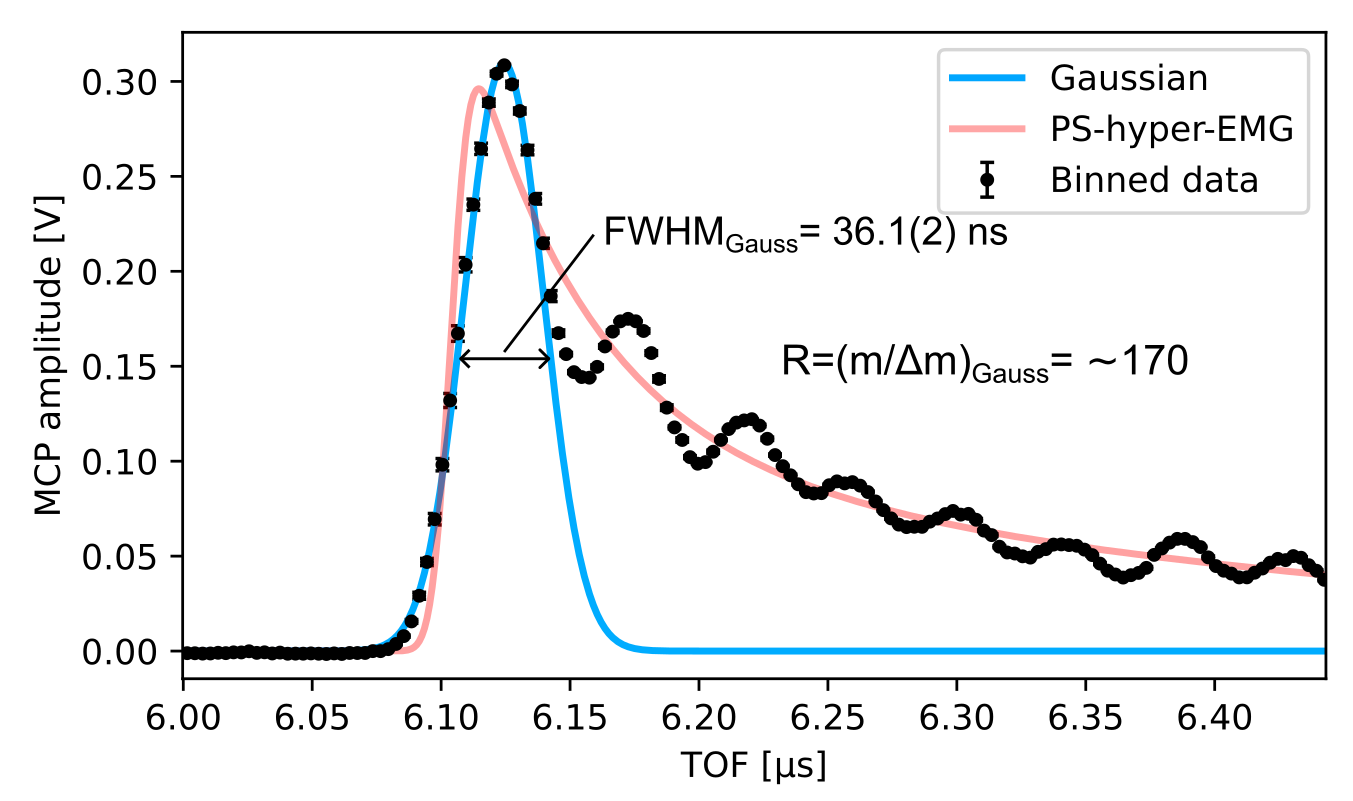}
\caption{TOF spectrum of ejected electron-cooled antiprotons ejected from a trap using the electrode bias $\mathrm{V_{launch}=-180}$ V and $\mathrm{V_{wall}}=-190$ V, fitted using PS-hyper-EMG function (red) and the Gaussian core (blue).}
\label{fig:calibration}
\end{figure}

\textit{Antiproton TOF calibration} In order to characterize the TOF spectrometry procedure, electron cooled antiprotons were prepared in a reverse polarity trap and ejected into the MCP (step 5 and 6 in Fig. \ref{fig:procedure}). Assuming a constant velocity approximation, the TOF ($T_{TOF}$) of the released ions is given by,  $T_{TOF} = L_{\text{eff}} \left( \frac{1}{2V}  \frac{m}{q}   \right)^{\frac{1}{2}}$ where $L_{\text{eff}}$ is the effective distance traveled assuming constant velocity, $m$ is the mass of the ion, $V$ is the accelerating voltage applied to the ion, and $q$ is the charge. The sample spectrum acquired using $\mathrm{V_{launch}=-180}$ V and $\mathrm{V_{wall}}=-190$ V, presented in Fig. \ref{fig:calibration}, is fitted using a double tailed PS-hyper-EMG distribution \cite{purushothaman2017hyper} and separately using a gaussian function fitted to the data in a window 60 ns around the peak. A TOF of $t_G =$\qty{6.1243(1)}{\micro\s}  with FWHM$_G = 36.1(2)\ \text{ns}$ gives a mass resolution of $R= \frac{t_0}{2 \Delta t_{\mathrm{FWHM}}}= M/\Delta M \approx 170$ for the gaussian core where $t_0$ is the peak centroid and $\Delta t_{\mathrm{FWHM}}$ the full width at half maximum of the fitted function. A mass resolution of $M/\Delta M\approx 45$ was obtained by numerically computing the FWHM of the fitted PS-hyper-EMG to account for the peak asymmetry. This observed asymmetry is partly a result of the spatial distribution of the antiprotons in the trap prior to the release, convoluted with the MCP detector response which had a slight impedance mismatching during the measurement.

Systematic studies were performed under different trap configurations to characterize the TOF ejection measurement for the ion studies. 

 The centroid $\mu$ represents the reference TOF of the antiprotons, which was used to calibrate the TOF spectra of other ions. In order to calibrate the TOF spectrum, the mass-to-charge ratio $m_i/q_i$ was calculated from,
\begin{equation}
\frac{m_i}{q_i} = \left( \frac{T_i}{T_{\text{ref}}} \right)^2 \frac{m_{\text{ref}}}{q_{\text{ref}}}, \label{eq:Calibration}
\end{equation}
where $T_i$ and $T_{\text{ref}}$ are the TOFs of the charged particle species of interest and the reference particle, such as the antiproton reference, respectively, where $\left(\tfrac{m}{q}\right)_{\bar{p}} = 1.007276\,\mathrm{u/e}.$ for the antiproton.

\subsection{Trap Configurations and Voltage Calibration}

\begin{figure}[t]
    \centering
    \includegraphics[width=\columnwidth]{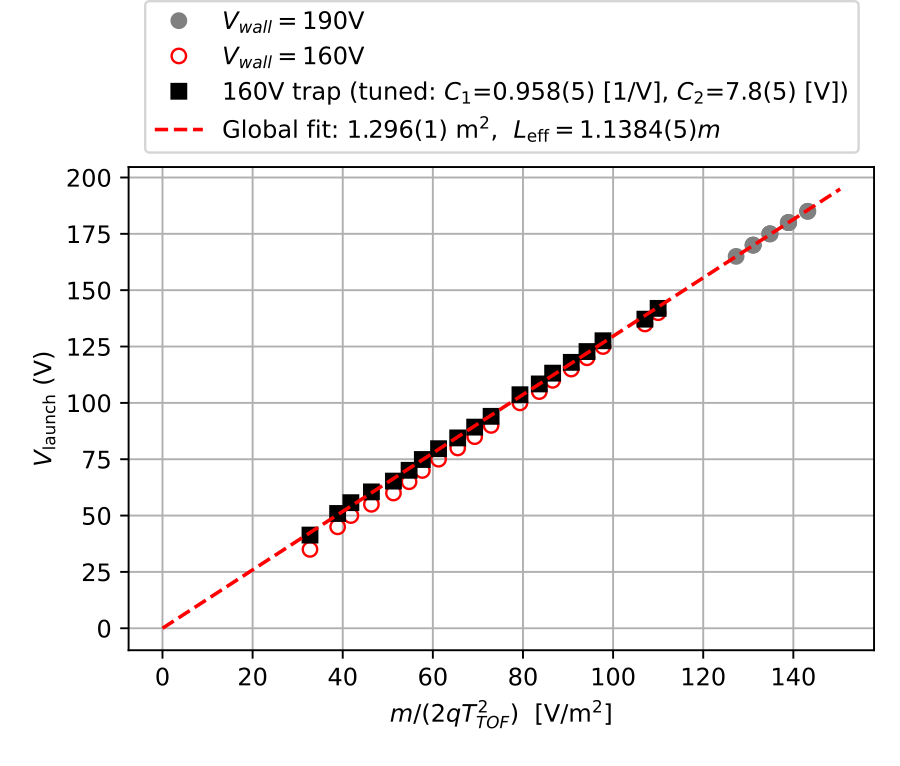}
    \caption{Calibration of the TOF spectrometer. Launch potentials $V_{\mathrm{set}}$ and corrected $V_{\mathrm{launch}}$ are plotted against $m/(2 q T_{\mathrm{TOF}}^2)$. The 190~V trap shows direct proportionality, while the 160~V trap requires correction using $V_{\mathrm{launch}} = C_1 V_{\mathrm{set}} + C_2$ with $C_1 = 0.958(5)$ and $C_2 = 7.8(5)\,\mathrm{V}$. The global fit yields $L_{\mathrm{eff}} = 1.1384(5)\,$m.}
    \label{fig:calibration_linear}
\end{figure}

TOF calibration measurements using cold antiprotons were performed with two trap configurations, $\mathrm{V_{wall}} = -190\,$V and $\mathrm{V_{wall}} = -160\,$V, while varying the applied launch potential $\mathrm{V_{launch}}$ on the central electrode. We study the characteristics of the TOF procedure by plotting $V_{\mathrm{launch}}$ against $m/(2 q T_{\mathrm{TOF}}^2)$ (see Fig.~\ref{fig:calibration_linear}). While the measurements performed using $\mathrm{V_{wall}} = -190\,$V are in good agreement with the proportionality expected from the constant-velocity assumption, the measurements using $\mathrm{V_{wall}} = -160\,$V deviate slightly. This deviation is expected since the set voltage $V^{\mathrm{set}}_{\mathrm{launch}}$ of the deeper trap differs from the actual potential experienced by the charged particles at the trap center, here denoted $V^{\mathrm{real}}_{\mathrm{launch}}$. The true launch potential is therefore parameterized as
\begin{equation}
    V^{\mathrm{real}}_{\mathrm{launch}} = C_1 V^{\mathrm{set}}_{\mathrm{launch}} + C_2,
\end{equation}
with $C_1$ and $C_2$ being calibration constants. These parameters could be determined, including their uncertainties, by minimizing the global $\chi^2$ of the combined data sets relative to the proportional linear relation $V^{\mathrm{real}}_{\mathrm{launch}} = k \,\frac{m}{2 q T_{\mathrm{TOF}}^2},$ where $k=L_{\mathrm{eff}}^2$ is the global fit coefficient. In this procedure, the $\mathrm{V_{wall}}=-190\,$V configuration was taken as a fixed reference, while $C_1$ and $C_2$ for the $\mathrm{V_{wall}}=-160\,$\textit{V} configuration were varied until the best agreement with the common fit was obtained. The extracted calibration constants are $C_1 = 0.958(5)$ (\textit{1/V}) and $C_2 = 7.8(5)$\textit{V}.  

This correction restores consistency across both configurations and enables a unified determination of the effective flight length. From the global corrected fit we extract an effective distance of $L_{\mathrm{eff}} = 1.1384(5)\,\mathrm{m}$, slightly larger than the measured geometric flight path of $1.05(1)\,\mathrm{m}$, that can be attributed to the constant-velocity approximation and the trap geometry.  

\subsection{Analysis of the Argon/Helium TOF Spectrum}

\subsection*{1. Data Preprocessing and Ringing Removal}
The raw TOF data from the four experimental runs were combined to create a single dataset. The resulting signal was binned into 70~ns intervals, and the mean voltage and standard error of the mean were calculated for each bin. As the MCP detector produces negative pulses, the spectrum was inverted to facilitate the analysis of positive peaks.

The intense \ce{He^1+} peak, arriving at approximately 12~$\mu$s, induced a significant electronic ringing artifact that manifested as a persistent oscillation across the spectrum. To isolate the smaller ion peaks for analysis, this artifact was modeled and subtracted. The model consisted of a composite function describing both the primary \ce{He^1+} peak and the ringing it produced. The peak itself was modeled with an Exponentially Modified Gaussian (EMG) distribution (defined in Eq.~A2), while the ringing was described by a sum of two distinct damped sinusoidal functions of the form:
\begin{equation*}
V_{\text{ring}}(t) = \sum_{i=1}^{2} C_i \exp\left(-\frac{t - t_{0,i}}{\tau_{r,i}}\right) \sin(2\pi f_i (t - t_{0,i}) + \phi_i)
\end{equation*}
where for each component $i$, $C$ is the amplitude, $\tau_r$ is the decay time, $f$ is the frequency, $\phi$ is the phase, and $t_0$ is the start time relative to the main peak. 

This full composite model was fitted to the binned spectrum using a least-squares algorithm. To ensure the model accurately captured the background artifact without being biased by the other ion peaks, the fit was constrained using time windows containing only the main \ce{He^1+} peak and representative regions of the oscillating background . After obtaining the best-fit parameters, the contribution from the two damped sinusoids was calculated and subtracted from the entire spectrum. The result of this procedure was a 'cleaned' spectrum with the ringing artifact removed, which was used for the subsequent multi-peak fit and calibration.

\subsection*{2. Multi-Peak Fit and TOF Determination}
The cleaned spectrum was fitted with a global model consisting of a sum of seven EMG functions, one for each identified ion species, and a constant background. This 29-parameter function was fitted to the data using a weighted least-squares algorithm, where the weights were the inverse square of the propagated standard errors from the binning stage. To ensure conservative error estimates, the resulting covariance matrix from the fit was scaled by the reduced chi-squared ($\chi^2_\nu$) of the fit if this value exceeded unity.

For an asymmetric EMG distribution, the peak maximum (the mode, $t_{\text{mode}}$) is shifted relative to the distribution's location parameter, $\mu$. Therefore, the final TOF for each ion was determined by numerically calculating the mode of its corresponding fitted EMG function. The uncertainty on this TOF value was calculated by analytically propagating the errors from the relevant parameters ($\mu, \sigma, \tau$) in the scaled covariance matrix.

\subsection*{3. Mass-to-Charge Calibration and Linearity Correction}
\begin{figure}[htbp]
    \centering
    \includegraphics[width=\columnwidth]{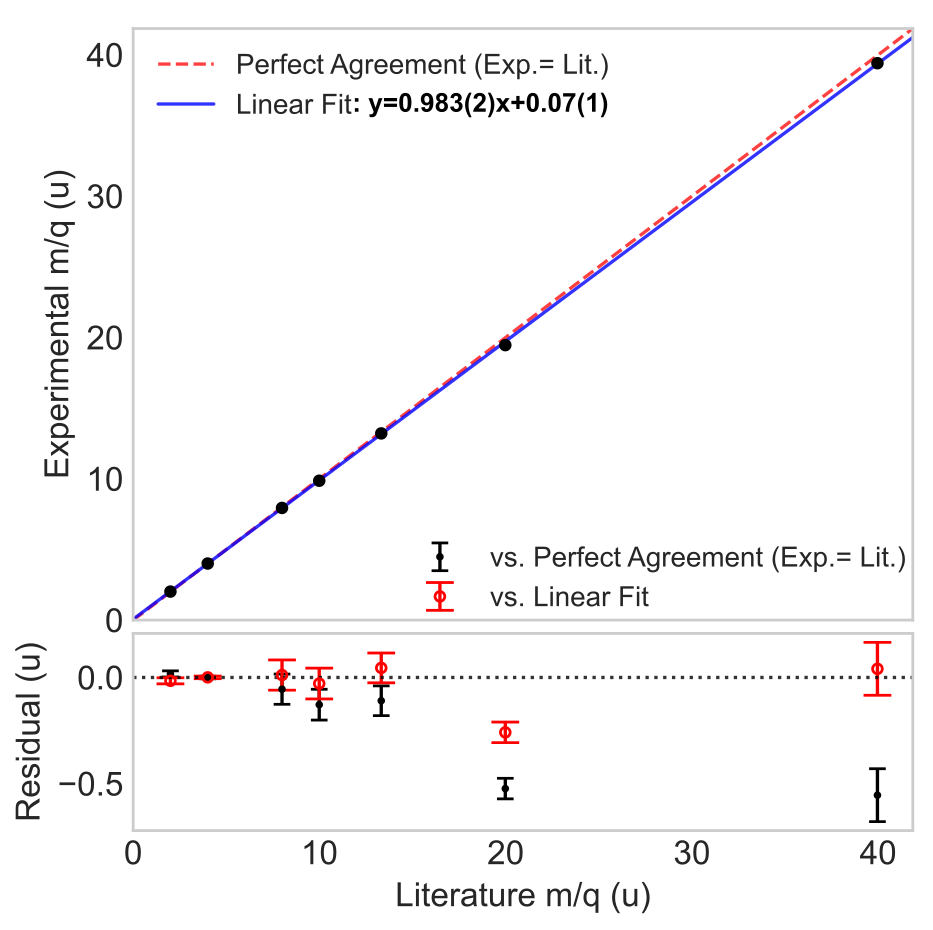}
    \caption{Calibration of the experimental mass-to-charge ($m/q$) axis. (Top) Experimental $m/q$ values, determined using a single-point calibration with \ce{He^1+}, are plotted against their literature values (black points). The red dashed line indicates a perfect 1:1 correlation, while the blue solid line shows a weighted linear regression fit to the data, which corrects for systematic instrumental effects. (Bottom) Residuals of the experimental data with respect to the ideal 1:1 line (black points) reveal a clear systematic trend, especially for heavier ions. Residuals with respect to the linear fit (red circles) are scattered around zero, demonstrating the effectiveness of the linear correction.}
    \label{fig:mq_calibration_fit}
\end{figure}

The TOF values were converted to a mass-to-charge ($m/q$) axis using Eq. \Ref{eq:Calibration}, with the precisely determined TOF mode of the \ce{He^1+} peak ($T_{\text{ref}}$) serving as a single-point reference. The statistical uncertainties on the calculated $m/q$ values, as reported in Table~I, were derived by propagating the uncertainties from the fitted TOF modes of the ion of interest ($\delta T_i$) and the reference ion ($\delta T_{\text{ref}}$) using the standard formula:
\begin{equation*}
\delta\left(\frac{m}{q}\right)_i = 2 \left(\frac{m}{q}\right)_i \sqrt{\left(\frac{\delta T_i}{T_i}\right)^2 + \left(\frac{\delta T_{\text{ref}}}{T_{\text{ref}}}\right)^2}
\end{equation*}
This single-point calibration assumes a perfectly quadratic relationship between TOF and $m/q$ over the full range. However, the data in Table~I presented in Fig \ref{fig:mq_calibration_fit} shows that this approximation works well for ions with $m/q$ values close to the \ce{He^1+} reference, but systematically deviates for the slower ions.

To characterize this deviation, a weighted linear regression was performed on the experimental $m/q$ values versus their known literature values, see Fig \ref{fig:mq_calibration_fit}. This fit provides a linear correction that can account for the observed offset and more accurately calibrate the mass axis across the entire detected range.

\bibliography{Bibliography}

\end{document}

%% file: Authorlist.tex
\author{F. P. Gustafsson}
\email{fredrik.parnefjord.gustafsson@cern.ch}
\affiliation{EP Department, CERN, 1211 Geneva 23, Switzerland}

\author{M. Volponi}
\affiliation{EP Department, CERN, 1211 Geneva 23, Switzerland}

\author{J. Zielinski}
\affiliation{Faculty of Physics, Warsaw University of Technology, ul. Koszykowa 75, 00-662 Warsaw, Poland}

\author{A. Asare}
\affiliation{University of Latvia, Department of Physics, Raina Boulevard 19, LV-1586, Riga, Latvia}

\author{I. Hwang}
\affiliation{Department of Physics, Yonsei University, Seoul, South Korea}

\author{S.~Alfaro~Campos}
\affiliation{University of Siegen, Department of Physics, Walter-Flex-Strasse 3, 57072 Siegen, Germany}
\affiliation{Universität Innsbruck, Institut für Experimentalphysik, Technikerstrasse 25/4, 6020 Innsbruck, Austria}

\author{M.~Auzins}
\affiliation{University of Latvia, Department of Physics, Raina Boulevard 19, LV-1586, Riga, Latvia}

\author{D.~Bhanushali}
\affiliation{University of Siegen, Department of Physics, Walter-Flex-Strasse 3, 57072 Siegen, Germany}

\author{A.~Bhartia}
\affiliation{University of Siegen, Department of Physics, Walter-Flex-Strasse 3, 57072 Siegen, Germany}
\affiliation{Université de Genève, 24 rue du Général-Dufour, 1211 Genève 4, Switzerland}

\author{M.~Berghold}
\affiliation{Heinz Maier Leibnitz Zentrum (MLZ), Technical University of Munich, Lichtenbergstraße 1, 85748, Garching, Germany}

\author{R. S. Brusa}
\affiliation{Department of Physics, University of Trento, via Sommarive 14, 38123 Povo, Trento, Italy}
\affiliation{TIFPA/INFN Trento, via Sommarive 14, 38123 Povo, Trento, Italy}

\author{K. Calik}
\affiliation{Faculty of Physics, Warsaw University of Technology, ul. Koszykowa 75, 00-662 Warsaw, Poland}

\author{A. Camper}
\affiliation{Department of Physics, University of Oslo, Sem Sælandsvei 24, 0371 Oslo, Norway}

\author{R. Caravita}
\affiliation{Department of Physics, University of Trento, via Sommarive 14, 38123 Povo, Trento, Italy}
\affiliation{TIFPA/INFN Trento, via Sommarive 14, 38123 Povo, Trento, Italy}

\author{F. Castelli}
\affiliation{INFN Milano, via Celoria 16, 20133 Milano, Italy}
\affiliation{Department of Physics ``Aldo Pontremoli'', University of Milano, via Celoria 16, 20133 Milano, Italy}

\author{G. Cerchiari}
\affiliation{University of Siegen, Department of Physics, Walter-Flex-Strasse 3, 57072 Siegen, Germany}
\affiliation{Universität Innsbruck, Institut für Experimentalphysik, Technikerstrasse 25/4, 6020 Innsbruck, Austria}

\author{S. Chandran}
\affiliation{Department of Physics, University of Liverpool, Liverpool L69 3BX, UK}

\author{A.~Chehaimi}
\affiliation{Department of Physics, University of Trento, via Sommarive 14, 38123 Povo, Trento, Italy}
\affiliation{TIFPA/INFN Trento, via Sommarive 14, 38123 Povo, Trento, Italy}

\author{S.~Choudapurkar}
\affiliation{University of Siegen, Department of Physics, Walter-Flex-Strasse 3, 57072 Siegen, Germany}

\author{R. Ciury\l{}o}
\affiliation{Institute of Physics, Faculty of Physics, Astronomy, and Informatics, Nicolaus Copernicus University in Torun, Grudziadzka 5, 87-100 Torun, Poland}

\author{P.~Conte}
\affiliation{INFN Milano, via Celoria 16, 20133 Milano, Italy}
\affiliation{Department of Aerospace Science and Technology, Politecnico di Milano, via La Masa 34, 20156 Milano, Italy}

\author{G. Consolati}
\affiliation{INFN Milano, via Celoria 16, 20133 Milano, Italy}
\affiliation{Department of Aerospace Science and Technology, Politecnico di Milano, via La Masa 34, 20156 Milano, Italy}

\author{M. Doser}
\affiliation{EP Department, CERN, 1211 Geneva 23, Switzerland}

\author{R.~Ferguson}
\affiliation{Department of Physics, University of Trento, via Sommarive 14, 38123 Povo, Trento, Italy}
\affiliation{TIFPA/INFN Trento, via Sommarive 14, 38123 Povo, Trento, Italy}

\author{M. Germann}
\affiliation{EP Department, CERN, 1211 Geneva 23, Switzerland}

\author{A. Giszczak}
\affiliation{Faculty of Physics, Warsaw University of Technology, ul. Koszykowa 75, 00-662 Warsaw, Poland}

\author{L.~T.~Gl\"{o}ggler}
\affiliation{EP Department, CERN, 1211 Geneva 23, Switzerland}

\author{\L.~Graczykowski}
\affiliation{Faculty of Physics, Warsaw University of Technology, ul. Koszykowa 75, 00-662 Warsaw, Poland}

\author{M. Grosbart}
\affiliation{EP Department, CERN, 1211 Geneva 23, Switzerland}

\author{F. Guatieri}
\affiliation{Department of Physics, University of Trento, via Sommarive 14, 38123 Povo, Trento, Italy}
\affiliation{TIFPA/INFN Trento, via Sommarive 14, 38123 Povo, Trento, Italy}

\author{N.~Gusakova}
\affiliation{EP Department, CERN, 1211 Geneva 23, Switzerland}
\affiliation{Department of Physics, University of Oslo, Sem Sælandsvei 24, 0371 Oslo, Norway}

\author{S. Haider}
\affiliation{EP Department, CERN, 1211 Geneva 23, Switzerland}

\author{S. Huck}
\affiliation{EP Department, CERN, 1211 Geneva 23, Switzerland}
\affiliation{Institute for Experimental Physics, Universität Hamburg, 22607 Hamburg, Germany}

\author{C.~Hugenschmidt}
\affiliation{Heinz Maier Leibnitz Zentrum (MLZ), Technical University of Munich, Lichtenbergstraße 1, 85748, Garching, Germany}

\author{M.~Jakubowska}
\affiliation{Faculty of Physics, Warsaw University of Technology, ul. Koszykowa 75, 00-662 Warsaw, Poland}

\author{M. A. Janik}
\affiliation{Faculty of Physics, Warsaw University of Technology, ul. Koszykowa 75, 00-662 Warsaw, Poland}

\author{G. Kasprowicz}
\affiliation{Faculty of Electronics and Information Technology, Warsaw University of Technology, ul. Nowowiejska 15/19, 00-665 Warsaw, Poland}

\author{K. Kempny}
\affiliation{Faculty of Physics, Warsaw University of Technology, ul. Koszykowa 75, 00-662 Warsaw, Poland}

\author{G.~Khatri}
\affiliation{EP Department, CERN, 1211 Geneva 23, Switzerland}

\author{A.~Kisiel}
\affiliation{Faculty of Physics, Warsaw University of Technology, ul. Koszykowa 75, 00-662 Warsaw, Poland}

\author{\L. K\l osowski}
\affiliation{Institute of Physics, Faculty of Physics, Astronomy, and Informatics, Nicolaus Copernicus University in Torun, Grudziadzka 5, 87-100 Torun, Poland}

\author{G. Kornakov}
\affiliation{Faculty of Physics, Warsaw University of Technology, ul. Koszykowa 75, 00-662 Warsaw, Poland}

\author{V. Krumins}
\affiliation{EP Department, CERN, 1211 Geneva 23, Switzerland}
\affiliation{University of Latvia, Department of Physics, Raina Boulevard 19, LV-1586, Riga, Latvia}

\author{L. Lappo}
\affiliation{Faculty of Physics, Warsaw University of Technology, ul. Koszykowa 75, 00-662 Warsaw, Poland}

\author{A. Linek}
\affiliation{Institute of Physics, Faculty of Physics, Astronomy, and Informatics, Nicolaus Copernicus University in Torun, Grudziadzka 5, 87-100 Torun, Poland}

\author{S.Mariazzi}
\affiliation{TIFPA/INFN Trento, via Sommarive 14, 38123 Povo, Trento, Italy}
\affiliation{Department of Physics, University of Trento, via Sommarive 14, 38123 Povo, Trento, Italy}

\author{P. Moskal}
\affiliation{Marian Smoluchowski Institute of Physics, Jagiellonian University, Krakow, Poland}
\affiliation{Centre for Theranostics, Jagiellonian University, Krakow, Poland}

\author{M.~M\"{u}nster}
\affiliation{Heinz Maier Leibnitz Zentrum (MLZ), Technical University of Munich, Lichtenbergstraße 1, 85748, Garching, Germany}

\author{P. Pandey}
\affiliation{Marian Smoluchowski Institute of Physics, Jagiellonian University, Krakow, Poland}
\affiliation{Centre for Theranostics, Jagiellonian University, Krakow, Poland}

\author{L. Penasa}
\affiliation{Department of Physics, University of Trento, via Sommarive 14, 38123 Povo, Trento, Italy}
\affiliation{TIFPA/INFN Trento, via Sommarive 14, 38123 Povo, Trento, Italy}

\author{M. Piwi\'nski}
\affiliation{Institute of Physics, Faculty of Physics, Astronomy, and Informatics, Nicolaus Copernicus University in Torun, Grudziadzka 5, 87-100 Torun, Poland}

\author{F. Prelz}
\affiliation{INFN Milano, via Celoria 16, 20133 Milano, Italy}

\author{T. Rauschendorfer}
\affiliation{EP Department, CERN, 1211 Geneva 23, Switzerland}
\affiliation{Department of Aerospace Science and Technology, Politecnico di Milano, via La Masa 34, 20156 Milano, Italy}

\author{B. S. Rawat}
\affiliation{Department of Physics, University of Liverpool, Liverpool L69 3BX, UK}
\affiliation{The Cockcroft Institute, Daresbury, Warrington WA4 4AD, UK}

\author{B. Rien\"{a}cker}
\affiliation{Department of Physics, University of Liverpool, Liverpool L69 3BX, UK}

\author{V. Rodin}
\affiliation{Department of Physics, University of Liverpool, Liverpool L69 3BX, UK}

\author{H. Sandaker}
\affiliation{Department of Physics, University of Oslo, Sem Sælandsvei 24, 0371 Oslo, Norway}

\author{S. Sharma}
\affiliation{Marian Smoluchowski Institute of Physics, Jagiellonian University, Krakow, Poland}
\affiliation{Centre for Theranostics, Jagiellonian University, Krakow, Poland}

\author{T. Sowi\'nski}
\affiliation{Institute of Physics, Polish Academy of Sciences, Aleja Lotnikow 32/46, PL-02668 Warsaw, Poland}

\author{E.~Tēberga}
\affiliation{University of Latvia, Department of Physics, Raina Boulevard 19, LV-1586, Riga, Latvia}

\author{M.~Tockner}
\affiliation{University of Siegen, Department of Physics, Walter-Flex-Strasse 3, 57072 Siegen, Germany}

\author{C. P. Welsch}
\affiliation{Department of Physics, University of Liverpool, Liverpool L69 3BX, UK}
\affiliation{The Cockcroft Institute, Daresbury, Warrington WA4 4AD, UK}

\author{M. Zawada}
\affiliation{Institute of Physics, Faculty of Physics, Astronomy, and Informatics, Nicolaus Copernicus University in Torun, Grudziadzka 5, 87-100 Torun, Poland}

\author{N. Zurlo}
\affiliation{INFN Pavia, via Bassi 6, 27100 Pavia, Italy}
\affiliation{Department of Civil, Environmental, Architectural Engineering and Mathematics, University of Brescia, via Branze 43, 25123 Brescia, Italy}